# SYSTEMATIC ERRORS IN GALILEO'S ASTRONOMICAL OBSERVATIONS AND ALLEGED ANOMALIES IN THE POSITION OF NEPTUNE

**Enrico Bernieri**
*INFN Sezione di Roma Tre and Dipartimento di Matematica e Fisica, Università Roma Tre, Rome, Italy.*
Email: bernieri@roma3.infn.it

**Gheorghe Stratan**
*Bogoliubov Laboratory of Theoretical Physics, JINR Dubna, Russian Federation and Horia Hulubei Institute of Physics and Nuclear Engineering, Bucharest, Romania.*

**Sara Bacchini**
*Dipartimento di Matematica e Fisica, Università Roma Tre, Rome, Italy.*

and

**Liviu Mircea**
*Astronomical Observatory of Roumanian Academy, Cluj-Napoca Branch, Romania.*

**Abstract:** In 1980 Kowal and Drake found that in December 1612 and January 1613 Galileo observed the planet Neptune. At that time, according to these authors, Galileo was able to measure angular separations with an accuracy of about 10 seconds of arc. However, as noticed by Kowal and Drake, the position of Neptune reported by Galileo is wrong with respect to the position computed with the modern ephemeris of about 1 minute of arc. This led Kowal and Drake to speculate on the possible errors of modern ephemeris of Neptune and sparked some debate about Neptune's ephemeris and/or possible errors in Galileo's measures. Until today, this anomaly has remained without a conclusive answer. Here we show that, in addition to the random errors, there are other significant measurement errors present in Galileo's observations. These errors may help clarify the origin of the alleged anomalies in the position of Neptune.

## 1 INTRODUCTION

In 1980 when analysing Galileo's manuscripts, Kowal and Drake (1980) showed that in December 1612 and January 1613, Galileo casually observed the planet Neptune—at that moment in close conjunction with Jupiter—which he marked as a fixed star, and he noted its motion 234 years before it was discovered. At that time, Galileo was mainly interested in the study of the periods of Jupiter's moons, discovered by him in 1610, and, even though he noticed that the strange star was moving and followed its movement for a few nights, he did not attach any particular importance to the observation. This probably also happened because, without an adequate telescope mounting it would have been impossible for him to follow up on this observation after Neptune—which was invisible to the naked eye—moved out of the field of view of his telescope, which he had centred on Jupiter.

According to Kowal and Drake (ibid.), at that time Galileo was able to measure angular separations with an accuracy of the order of 10 arcsec. In the observations concerning Neptune of 26/27 and 27/28 January 1613 shown in Figure 1, Galileo draws a linear scale showing 24 Jovian radii. Assuming that the whole drawing is to scale, it is possible to measure the angular separation between Neptune and the star SAO 119234. The separation reported by Galileo is of about 3.75 Jovian radii, or 75 arcsec. However, the ephemeris shows that Neptune should have been about 130 arcsec from the star at this time. According to Kowal and Drake, this is a much greater inaccuracy than the typical Galilean accuracy in angular measurements.

This observation led Kowal and Drake to hypothesise that the modern ephemeris of Neptune is in error by a significant amount, which would require a revision of the orbital elements of the planet, and suggest the existence of an unknown perturbation. To support their hypothesis, the authors also reported a subsequent 'pre-discovery' measurement of Neptune's position, made by Lalande in 1795, that differs from the predicted

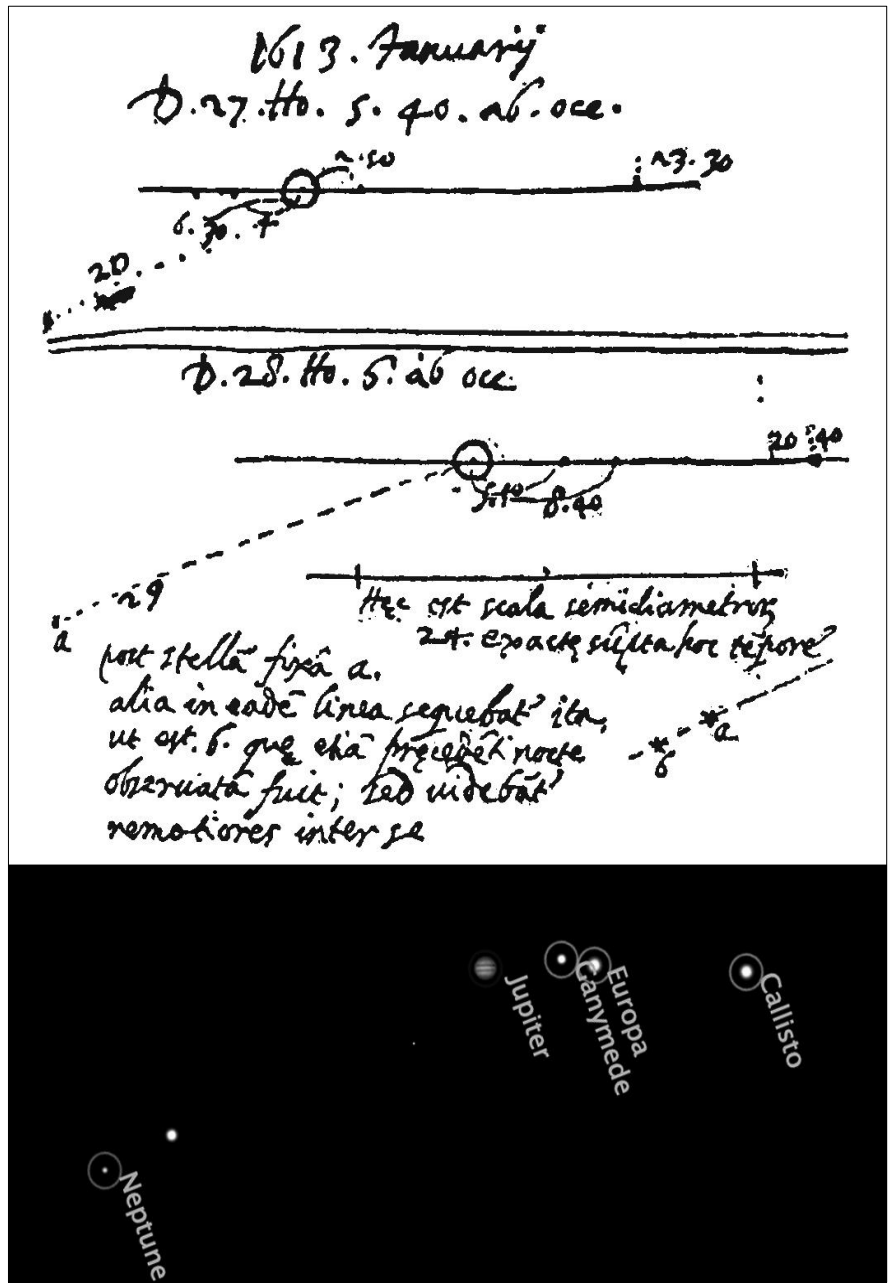

Figure 1 (Top): Galileo's notebook for his 26/27 (top drawing) and 27/28 (bottom drawing) January 1613 observations (after Kowal and Drake, 1980). In the lower drawing, it is possible to see Jupiter (circle) and the dashed line on the left showing the separation between Jupiter and the star 119234 in the Smithsonian Astrophysical Observatory (SAO) Star Catalogue. On the separated dashed line in the lower right corner, the letter 'b' indicates the planet Neptune and the letter 'a' the same star. The solid line under Jupiter indicates the scale of the drawing and corresponds to 24 Jovian radii. Bottom: the screenshot from Stellarium of the same observation.

position by 7 arcsec.

This hypothesis was questioned by Rawlins (1981) and Standish (1981), who carried out accurate calculations and showed that, even allowing for important perturbations of Neptune's orbit, it was extremely difficult to take account of the observed anomaly, because the 'incorrect' ephemeris position lay on the same straight line from Jupiter to the star.

In a subsequent paper, Standish and Nobili (1997) also tested the hypothesis of an unknown planet perturbing Neptune's orbit. However, their calculations showed that it would need a planet of 20 or more Earth masses to account for the effect observed. They concluded that if Neptune were perturbed to where Galileo drew it, the perturbation was a rare one—and pathological in nature.

Another obvious hypothesis is that there are mistakes in Galileo's drawings, or that the scale of the drawing is wrong, or that Galileo's measures are not sufficiently precise (Rawlins, 1981; Standish, 1981).

We will show that a further explanation is possible, which also explains other anomalies present in Galileo's observations and helps us to better understand the alleged anomaly in the position of Neptune.

## 2 GALILEO'S DATA

Galileo carried out a significant number of observations of the satellites of Jupiter, between 7 January 1610 and 19 November 1619. During this period, the quality of his telescopes and his observation techniques underwent notable evolution and improvements.

The most important improvement was the implementation of a kind of external micrometer[1] mounted to enhance the precision of his angular measurements, which Galileo started using on 31 January 1612 (Favaro, 1907a: 415, 446). With this instrument, he measured Jupiter's diameter, and used half this value as a unit to determine the elongations of the Jovian satellites from the planet's center. The entire procedure is described by Shea and Bascelli (2009).

In order to appreciate the Neptune-Jovian conjuncture, we decided to examine the precision of Galileo's observations covering the period from 1 March to 8 May 1613 and reported in his *Letter on Sunspots* (see Favaro, 1907b).

In that work, the observations are reported only in graphic form. Therefore, to avoid introducing further measurement errors, we used the same observations reported in his notebooks also in numerical form, in units of Jovian radii (Favaro, 1907a: 585–597). The reason of this choice is connected with the compactness in time of the observations (no gaps) and also to the fact that both the observation of Neptune and the observations analysed by us, are subsequent to the introduction of the micrometer and are not temporally very distant from each other, so it can be assumed that the precision of the measurements has not changed significantly over a period of about seventeen months.

From the above-mentioned period, we selected 119 observations[2] corresponding to 411 measurements of the elongations of Jovian satellites (their apparent angular distances from Jupiter and in some cases the angular distances between satellites) given in Jovian semi-diameters. To pass from semi-diameters, to angular separation between Jupiter and its satellites, we have to take into account the variation in time of Jovian apparent diameter, as seen from the Earth. For the accounted period, the average value of the apparent semi-diameter is 20 arcsec, with a total variation of about 1 arcsec.

To establish the accuracy of Galileo's observations, we compared his data with the those obtained by using modern planetary software, Stellarium (https://stellarium.org), an open-source planetarium that includes an 'angle measure' tool that allows for precise angular measurements.

As shown in a previous paper (Bernieri, 2012), the calculation algorithms used in modern planetary software[3] give a precision of around 1 arcsec or less, allowing for a reconstruction, with great reliability, of the 'true' positions of Jupiter's satellites at the times of Galileo's observations.

Regarding the time, previous analysis shows that the times Galileo cited were often exact and can at least be trusted to within 15 minutes (Drake and Kowal, 1980). In most cases, such a small error in the recorded time would not significantly affect the position of a satellite, implying an error of a few arc-seconds only in the position of Io, the inner-most (and fastest-moving) of Jupiter's satellites.

By subtracting from the Galileo data the 'true' values obtained with the ephemeris, it is possible to associate errors with each data point.

## 3 DATA ANALYSIS

Figure 2 shows the distribution of the errors. It is possible to notice that some errors are quite significant (of the order of about 100 arcsec) and that the distribution is strongly asymmetrical, with a prevalence of negative errors. In the presence of random errors only, it would have been expected, instead, that the distribution would be symmetric with a typically Gaussian shape. This shows that Galileo tended to systematically underestimate the angular separations.

Looking for the origin of this asymmetry, we plotted the errors as a function of the elongations, always obtained with modern ephemerides (Figure 3). The correlation between the two variables is quite evident,[4] showing that the error increases as the angular separation widens.

Figure 4 shows the distribution of random errors, obtained by subtracting from the errors the systematic trend, assumed to be described by a linear least-square fit. As we can see, the distribution is now symmetrical and is well fitted by a Gaussian curve, typical of random errors.[5]

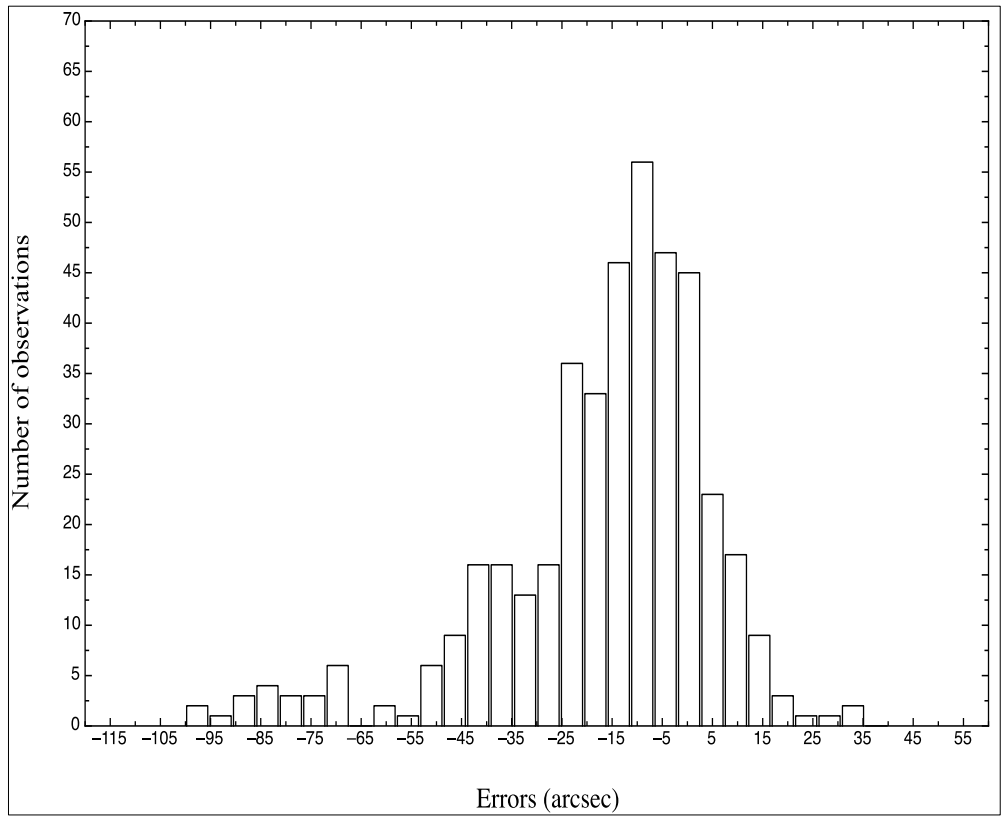

Figure 2. Distribution of Galileo's experimental errors. The distribution is strongly asymmetrical, with prevailing negative errors.

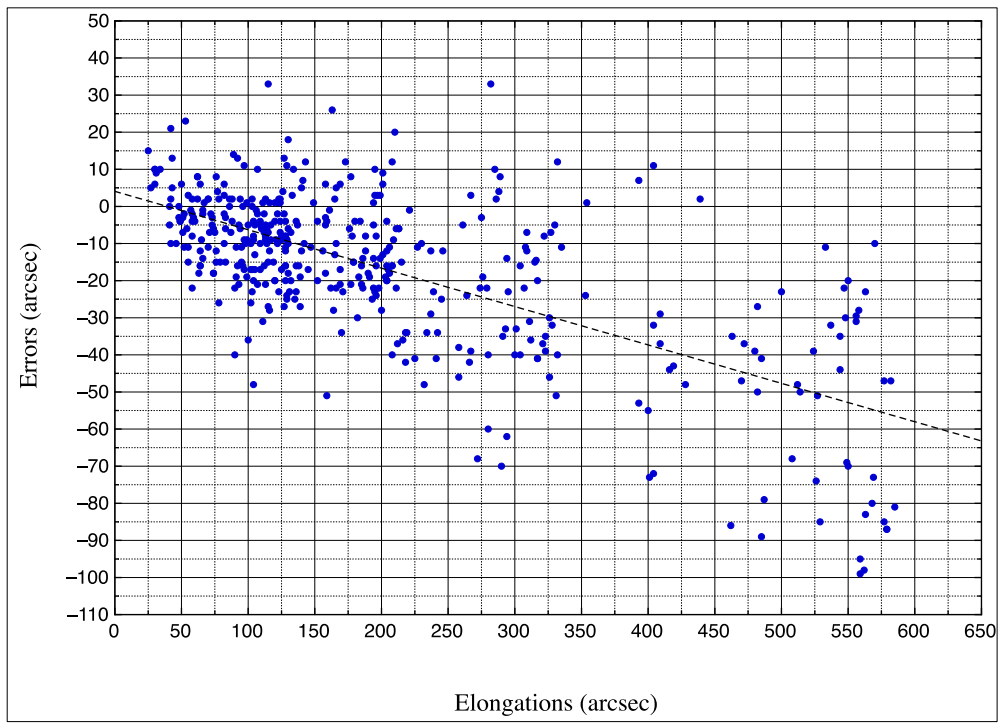

Figure 3: Plot of Galileo's errors in angular measurements as a function of the 'true' Jupiter satellite elongations, obtained by modern ephemeris. The line is a least-square fit of the data.

The standard deviation of the distribution is 10.8 arcsec, a value that is extremely close to the accuracy attributed by Kowal and Drake to Galileo's measurements. However, in light of our analysis, this value must be considered as the 'precision' (the degree to which repeated measurements under unchanged conditions show the same results) of Galileo's measurements and not as their 'accuracy' (the proximity of the measurement result to the true value), which is much worse because of the bias observed. It can therefore be said that, according to Kowal and Drake, Galileo's measurements are quite precise, but, according to our results, not as accurate.

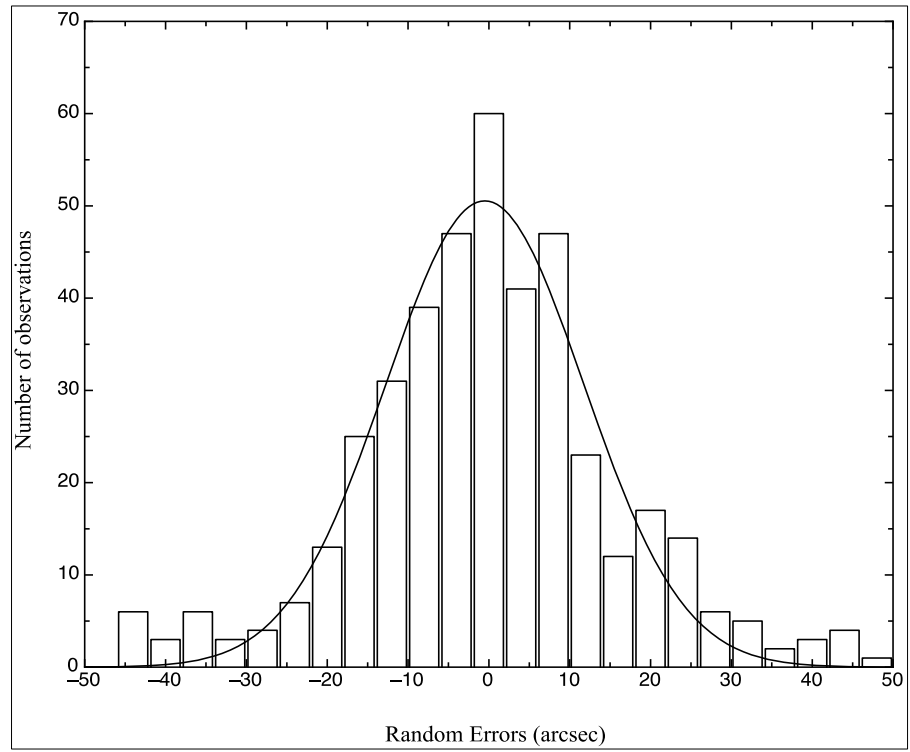

Figure 4: Distribution of Galileo's random errors, obtained by subtracting the systematic trend (shown in Figure 3) from the errors (shown in Figure 2). The line shows a Gaussian best fit of the data, with a standard deviation of 10.8 arcsec.

In our opinion, this effect has never been noticed before, as the random errors can mask the effect if the number of data analysed is not big enough.

As far as we know, the only other work concerning the precision and accuracy of Galileo's measurements is that of Graney (2007) which showed how over time the precision and resolution of Galileo's observations improved up to the diffraction limits of its telescopes. However, Graney's work analyses only a few particular cases and does not analyse statistically enough measurements to highlight systematic errors.

Incidentally, our measure of the precision of Galileo's measurements from 1613 data, shows that this is significantly improved over the *Sidereus Nuncius* measurements (1610) when the standard deviation of the statistical error was about 1 arcminute (Bernieri, 2012).

## 4 DISCUSSION

The systematic effect observed explains quite well some anomalies in the observation reported by Kowal and Drake. In fact, in the observation of 28 January, even the star SAO 119234 is in a wrong position being reported by Galileo at about 50 arcsec less than the value computed by the ephemeris, and the same star, on the previous observation of 2 January, was reported by Galileo as at 48 Jovian radii, instead of 52 Jovian radii, which is about 80 arcsec in defect. This effect may also explain quite well the wrong separation, smaller by about 100 arcsec, between Jupiter and Neptune. Looking at Figure 3, and considering that in the observation reported by Kowal and Drake (1980) the true angular elongation of Neptune from Jupiter was ~730 arcsec, it is easy to see—extrapolating the correlation line—that the linear trend is in good agreement with the observed error of about 100 arcsec.

However, this bias does not fully quantitatively explain the wrong separation, of 55 arcsec, between Neptune and SAO 119234, whose angular separation is 130 arcsec. As it can be seen from the trend line in Figure 3, at this elongation Galileo's systematic bias amounts to only 10 arcsec and this falls well short of accounting for the 55 arcsec error.

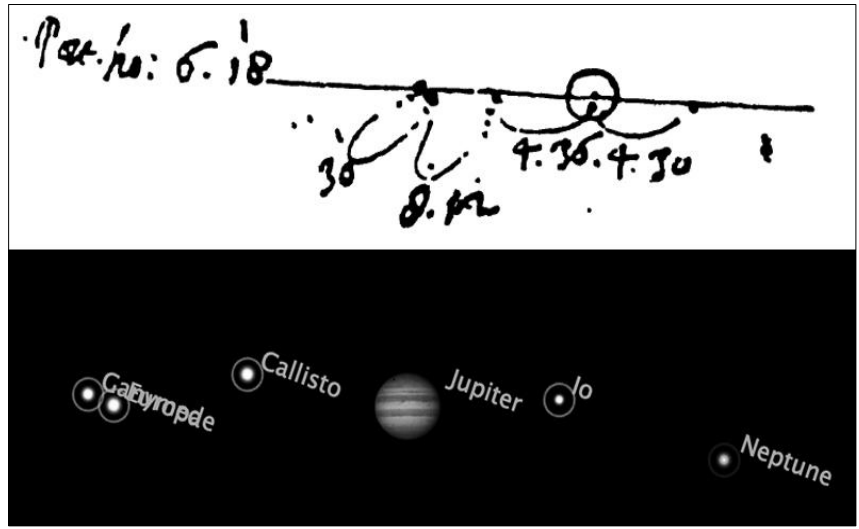

Figure 5 (Top): Detail of Galileo's drawing of 6 January 1613. In the opinion of Standish and Nobili (1997) the spot on the right is the planet Neptune. Bottom: A screenshot from Stellarium of the same observation.

Nevertheless, in the light of our analysis, this error may be due to a combination of systematic error and a particularly large random error. In any case, the 'wrong' position of Neptune cannot be seen only in light of the random errors (Galileo's measurement precision), according to Kowal and Drake, but also considering the systematic error observed (Galileo's measurement accuracy). It should also be noted that the elongation of the star SAO 119234 from Jupiter, as might be expected, is significantly wrong in defect.

Assuming that Galileo most likely placed Jupiter at the centre or near the centre of the field of view (FOV), this trend could show that the systematic error increases moving away from the centre of FOV.

Consequently, the accuracy of the measurements made by Galileo towards the edges of the FOV must be considered with great caution. This is the case of Neptune and SAO 119234, which were probably on the edge of the FOV, and relying on a single measure of this type to draw general conclusions can be very unreliable.

To confirm further the presence of a systematic error in the position of Neptune, we have analysed another possible observation of this planet, discovered by Standish and Nobili (1997) in a Galileo's drawing of 6 January 1613. Figure 5 shows a detail of this drawing, where the spot to the right appears to be the planet Neptune. By measuring the elongation between Jupiter and Neptune with a ruler we obtain a separation of about 7.7 Jovian radii, about 146 arcsec, while the value provided by the software is about 190 arcsec, corresponding to about 9.7 Jovian radii. Also, in this case, therefore, there is an error in defect (of about 44 arcsec) in the position of Neptune.

To seek an explanation for this systematic error, we have considered two possible causes:

(1) A change in magnification through Galileo's telescopic field of view, caused by a barrel distortion; and

(2) A wrong measurement of the Jovian radius, which Galileo used as a unit of measurement for elongations.

In the case of barrel distortion, the magnification decreases towards the edge of the field, and this fact could justify the errors in default made by Galileo, bigger and bigger as one moves away from the optical axis. It is well-known (e.g. see Jenkins and White, 1957), that this kind of barrel distortion arises when a stop is placed in front of a lens. This is exactly the configuration used by Galileo, who placed a narrow diaphragm in front of the lens of his telescope to reduce aberrations (Greco et al., 1993; 1992).

We performed some ray-tracing simulations using the optical configuration of the two Galilean telescopes now at the Science Museum of Florence (Greco et al. 1993). These simulations showed a certain amount of barrel distortion, which, however, was too small to fully explain the magnitude of the observed error.

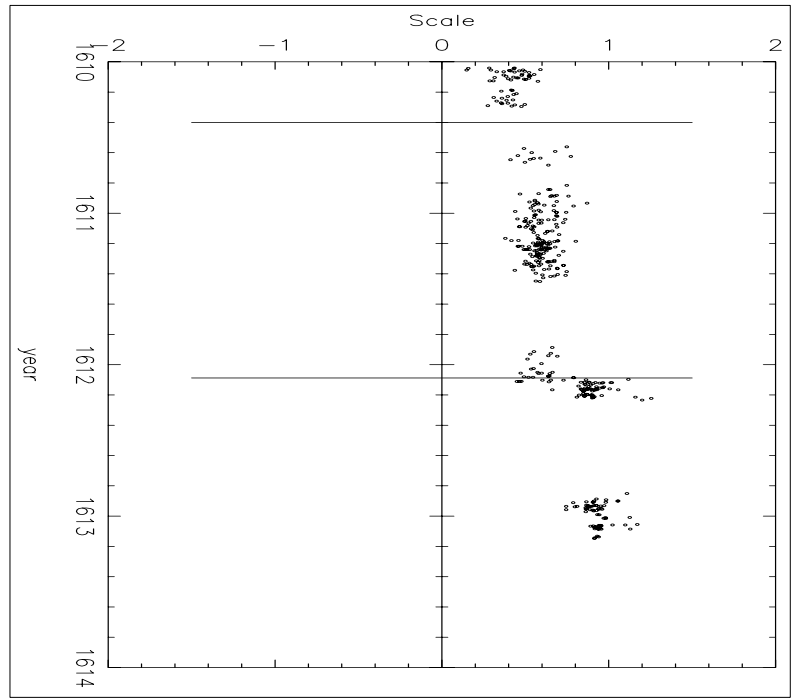

Figure 6: Plot of the scale factors vs. time for Galileo's observations over the four years, 1610–1613 (courtesy: M. Standish).

Regarding Jovian radius error, this hypothesis was first advanced by Myles Standish (M. Standish, pers. comm., March 2022).

It is quite evident, analysing the resolution, that the performance of Galileo's first telescopes was quite far from being diffraction-limited, and that Galileo overestimated the diameter of Jupiter (Bernieri, 2012). Subsequently, the optical characteristics of the lenses have improved,[6] and the introduction of the micrometer certainly improved the precision of the measurements. However, a combination of atmospheric seeing, the limited quality of his telescope, and the quality of the measuring device he used at that time, can, in our opinion, easily justify Galileos' systematic effect we observed. An overestimation of 0.1 Jovian radius ($R_j$) would be sufficient to justify the systematic error observed by us and this corresponds, for the measurements that we have analyzed, to an error of about 2 arcseconds, compatible with the causes indicated above.

This hypothesis is strongly justified by the analysis of two graphs produced by Myles Standish (pers. comm., January 2022)

Figure 6 plots the scale factors vs. time for Galileo's observations over the four years, 1610–1613. The scale factor is the ratio of Jupiter's actual radius to the radius measured by Galileo. One will see three distinct time intervals: January 1610–March 1610, July 1610–January 1612, and 16–12 February–March 1613. Evidently, from one of those intervals to the next, Galileo made some improvement to either his telescope or to the device he used for measuring the separations. Incidentally, it is interesting to observe the sharp improvement in the scale factor at the beginning of 1612, when Galileo began to use his micrometer. As time went on, the scale factor got closer to unity, showing improvement of the observations. However, from Figure 6 it can be seen that even in the last time interval (1612–1613) the scale factor is slightly lower than unity.

Figure 7 shows two of Galileo's 'Jovilabes' corresponding to the second- and third-time intervals shown in Figure 6. Jovilabes are, effectively, analog computers, where Galileo would compute the longitude for a satellite, use it to locate the satellite in its orbit on the Jovilabe, and then drop that position onto the x-axis, thus showing the elongation from Jupiter. Of interest here, however, are the sizes of the satellite orbits that he drew in terms of $R_j$. From the second Jiovilable (time interval 1612–1613) it can be observed that Ganymede and Callisto, whose correct semi-major axes are 15.0 and 26.3 $R_j$ respectively, has the semi-major axes of 14 and 24 SJ—too small by a ratio a little greater than 0.9, according to the magnitude of the effect we observed. In conclusion, the error that Galileo made in estimating the radius of Jupiter seems to be the major reason for the explanation of the systematic error that we found.

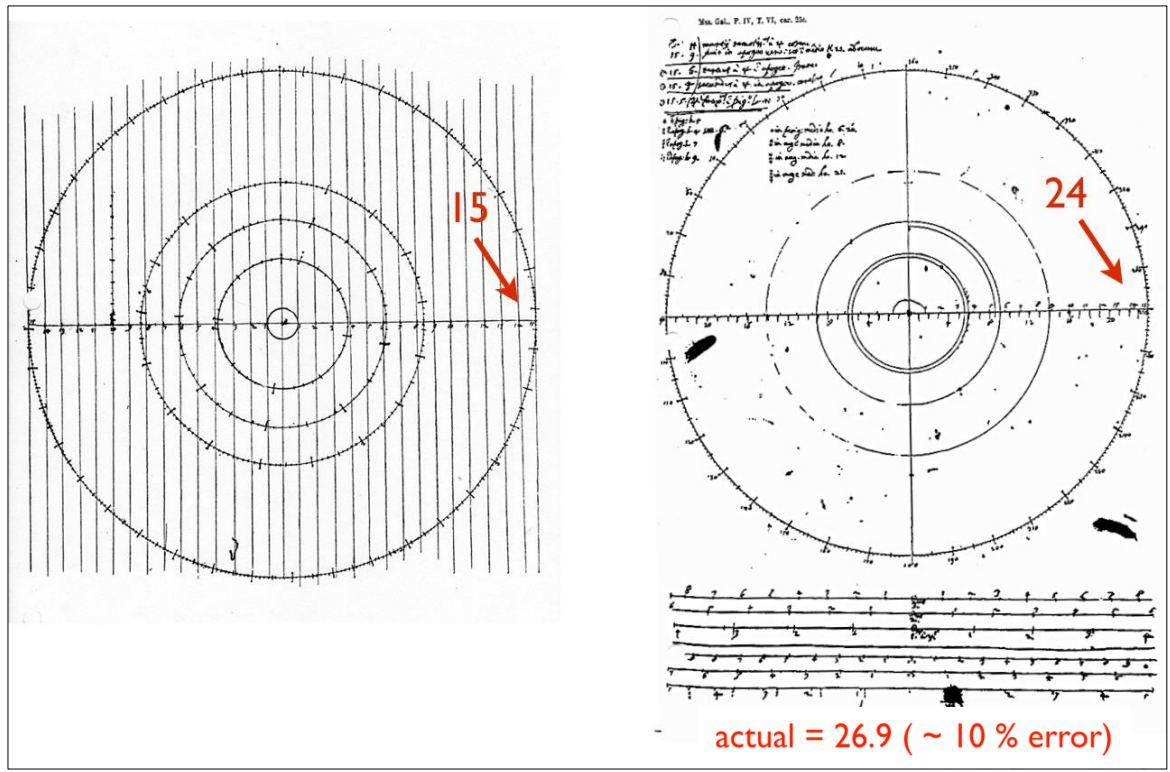

Figure 7: Two of Galileo's 'Jovilabes' corresponding to the second and third time intervals shown in Figure 6. For Ganymede and Callisto, whose correct semi-major axes are 15.0 and 26.3 SJ, respectively, the first Jovilabe gives 8.4 and 15.0 –too small by a ratio of nearly 0.6. The second Jovilabe (1612–1613) has the semi-major axes of Ganymede and Callisto being 14 and 24 SJ—too small by a ratio a bit greater than 0.9, in agreement with the data shown in Figure 3 (courtesy: M. Standish).

## 5 CONCLUSIONS

In this work, we performed an in-depth analysis of Galileo's measurement errors and showed that were not only of a random character but also that important systematic effects must be considered in analysing his observations. These effects, which have never been considered previously, help explain the lack of accuracy in many of the observations and provide, in our opinion, a solid indication that the alleged anomalies in the position of Neptune are due to measurement errors.

The origin of the systematic errors in Galileo's measurements is probably due to an overestimation of the radius of Jupiter, which Galileo used as a unit of measurement, although a contribution due to field distortion is probably present.

We plan to extend our analysis to a larger sample of Galileo's measurements, considering the entire period from 1610 to 1619 in which Galileo made astronomical observations, in order to obtain a more complete picture of the evolution over time of their precision and accuracy.

## 6 NOTES

1. Galileo's micrometer was described for the first time by his disciple Giovanni Alfonso Borelli in 1666 (Favaro, 1907: 415 and 447).
2. A small number of Galileo's measurements concerning this period are illegible or marked only by the symbols adopted for the satellites, and they were omitted from our investigation.
3. The routines can be found in the book *Astronomical Algorithms* (Meeus 2004), which sets out the theory elaborated by the astronomer J.H. Lieske, with improvements published in 1987 known as 'E2x3' (Lieske 1987).
4. Assuming a linear correlation and performing a linear best fit, we obtained a chi-squared of 377.7 and a corresponding goodness of fit of 0.92.
5. With a chi-squared of 15.75 and a goodness of fit of 0.87.
6. The performances of the telescopes preserved in the Science Museum of Florence appear to be diffraction limited (see Greco et al. 1992; 1993).

## 7  ACKNOWLEDGMENTS

We would like to thank very much Myles Standish for his fundamental contribution in the interpretation of our results and for the helpful discussions.

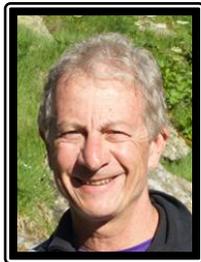

**Dr Enrico Bernieri** is a researcher at the National Institute of Nuclear Physics (INFN) and lecturer at the Department of Mathematics and Physics at the Roma Tre University. He is an experimental physicist specializing in detectors and data analysis and his scientific interests span many fields of physics and astronomy.

He worked in the field of synchrotron radiation, participated in experiments in the field of high energy physics and astrophysics and contributed to the drafting of the first catalog of high energy gamma sources observed by the Fermi - LAT space telescope.

He approached the study of the history of Astronomy, and in particular of Galileo's astronomical observations, during the International Year of Astronomy (2009), collaborating with Professor Gheorghe Stratan in the preparation of an exhibition relating to Galileo's observations reported in the *Sidereus Nuncius*. Since then his interest in this field has expanded and currently concerns the study of the evolution of the instruments and observational techniques developed over time by Galileo.

He currently teaches experimental astrophysics in the Department of Mathematics and Physics at the Roma Tre University, is in charge of the Department's teaching astronomical observatory and is involved in numerous educational and outreach projects in the field of astronomy.

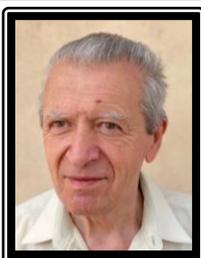

**Professor Gheorghe Stratan** was born in Romania in 1941. His activity started with the National Institute of Physics and Nuclear Engineering in Bucharest, in the field of Nuclear Theory, then in the Joint Institute for Nuclear Physics (JINR) - Dubna (RF), with the Department of Theoretical Physics.

In 2003, he won the concourse of Head, Chair of History of Science, inaugurated by him at Babes-Bolyai University in Cluj-Napoca. He is currently Senior Scientist I at JINR Dubna and has been a member of the Scientific Council of this institution from more than 20 years.

Professor Stratan translated into Romanian 12 books of famous scientists, both contemporary and past, many of which he commented on and updated. He prepared three international History of Science exhibitions (two of them about astronomy). His exhibition about Galileo's astronomical discoveries was presented to the public 18 times, in 12 cities in six countries and on three continents (the Italian version of this exhibition was co-authored with Professor Enrico Bernieri) .

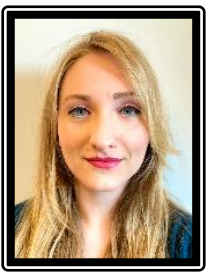

For his papers and books in Theoretical Nuclear Physics, History of Science and popularization of Science, Professor Stratan has received Romanian and international prizes and medals.

**Dr Sara Bacchini** was born in Rome in 1995 and completed a Bachelor's degree thesis on "Astronomical Imaging: Scientific and Aesthetic Contents" through the University of Roma Tre, supervised by Professor Enrico Bernieri.

In her student days, she was involved in Galileo's telescope studies and scientific education events. She has a long-standing experience in ray-tracing astronomy, and indeed in 2021 she concluded a Master's thesis on "Estimate of Wind Fields in the Jupiter Polar Regions From JIRAM-Juno Images", supervised by Professor Enzo Pascale, Dr Davide Grassi and Dr Giuseppe Piccioni.

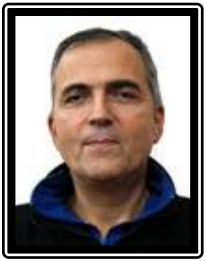

**Dr Liviu Mircea** was born in Romania in 1957. He has degrees in Mechanical Engineering (1982, Technical University of Cluj-Napoca) and Physics (1990, Babes-Bolyai University of Cluj-Napoca).

Since 1999 he has been a senior researcher at the Astronomical Observatory of Romanian Academy, Cluj-Napoca Branch. Its main activities and responsibilities cover many fields: data acquisition, observational astronomy in the field of artificial satellites, maintenance and adjustment of optical instruments and apparatus, and also observational astronomy and astrophysics (optical domain, CCD, photometry, astrometry, artificial satellites tracking, variable stars).

He also made important contributions in digitization of the Plate Archives of Astronomical Observatory of Cluj-Napoca, and in History of Astronomy and Astronomy popularization.